\documentclass[amssymb,useAMS,prd,aps,amsmath,nofootinbib,superscriptaddress,
twocolumn]{revtex4-1}
\usepackage[utf8]{inputenc}
\usepackage{natbib}
\usepackage{color}
\usepackage{graphicx}
\usepackage{amsmath}
\usepackage[caption=false]{subfig}
\usepackage[colorlinks=true,linkcolor=blue,citecolor=blue,urlcolor=black]{
hyperref}

\begin{document}
\title{The Significance of the 750 GeV Fluctuation in the ATLAS Run 2 Diphoton Data}
\author{Jonathan H. Davis}
\author{Malcolm Fairbairn}
\author{John Heal}
\author{Patrick Tunney}
\affiliation{Theoretical Particle Physics and Cosmology, Department of Physics, Kings College London, London WC2R 2LS, United Kingdom
\\ {\smallskip \tt  \href{mailto:jonathan.davis@kcl.ac.uk}{jonathan.davis@kcl.ac.uk},\href{mailto:malcolm.fairbairn@kcl.ac.uk}{malcolm.fairbairn@kcl.ac.uk},\href{mailto:john.heal@kcl.ac.uk}{john.heal@kcl.ac.uk},\href{mailto:patrick.tunney@kcl.ac.uk}{patrick.tunney@kcl.ac.uk}\smallskip}}
\date{11th January 2016}

\begin{abstract}
We investigate the robustness of the resonance like feature centred at around a 750~GeV invariant mass in the 13 TeV diphoton data, recently released by the ATLAS collaboration. We focus on the choice of empirical function used to model the 
continuum diphoton background in order to quantify the uncertainties in the analysis due to this choice. We extend the function chosen by the ATLAS collaboration to one with two components. By performing a profile likelihood analysis we find that the local significance of a resonance drops from $3.9\sigma$ using the ATLAS background function, and a freely-varying width, to only $2\sigma$ with our own function. We argue that the latter significance is more realistic, since the former was derived using a function which is fit almost entirely to the low-energy data, while underfitting in the region around the resonance.
\end{abstract}

\maketitle

\section{Introduction}
Following the discovery of the (Brout-Englert-)Higgs boson~\cite{Chatrchyan:2012xdj,Aad:2012tfa}, a central aim of the Large Hadron Collider (LHC) at CERN is to search for new particles beyond the Standard Model of particle physics. Such particles could show up in any one or more of the processes studied by the LHC experiments.  The LHC has recently released its first data from proton collisions at a centre-of-mass energy of $\sqrt{s} = 13$~TeV.  The most intriguing result of this recent data-release has been the potential observation of a resonance-like feature above the expected continuum background in the diphoton channel, at an invariant mass $m_{\gamma \gamma} \sim 750$~GeV, which has provoked a great deal of excitement amongst the theoretical community \cite{Harigaya:2015ezk,Mambrini:2015wyu,Backovic:2015fnp,Angelescu:2015uiz,Knapen:2015dap,Buttazzo:2015txu,Pilaftsis:2015ycr,Franceschini:2015kwy,DiChiara:2015vdm,Higaki:2015jag,McDermott:2015sck,Ellis:2015oso,Low:2015qep,Bellazzini:2015nxw,Gupta:2015zzs,Petersson:2015mkr,Molinaro:2015cwg,Chao:2015ttq,Fichet:2015vvy,Curtin:2015jcv,Bian:2015kjt,Chakrabortty:2015hff,Ahmed:2015uqt,Agrawal:2015dbf,Csaki:2015vek,Falkowski:2015swt,Aloni:2015mxa,Bai:2015nbs,Dutta:2015wqh,Cao:2015pto,Matsuzaki:2015che,Kobakhidze:2015ldh,Martinez:2015kmn,Cox:2015ckc,Becirevic:2015fmu,No:2015bsn,Demidov:2015zqn,Gabrielli:2015dhk,Benbrik:2015fyz,Kim:2015ron,Alves:2015jgx,Megias:2015ory,Carpenter:2015ucu,Bernon:2015abk,Chakraborty:2015jvs,Ding:2015rxx,Han:2015dlp,Han:2015qqj,Luo:2015yio,Chang:2015sdy,Bardhan:2015hcr,Feng:2015wil,Antipin:2015kgh,Wang:2015kuj,Cao:2015twy,Huang:2015evq,Liao:2015tow,Heckman:2015kqk,Dhuria:2015ufo,Bi:2015uqd,Kim:2015ksf,Berthier:2015vbb,Cho:2015nxy,Cline:2015msi,Bauer:2015boy,Chala:2015cev,Barducci:2015gtd,Kulkarni:2015gzu,Chao:2015nsm,Arun:2015ubr,Han:2015cty,Chang:2015bzc,Boucenna:2015pav,Murphy:2015kag,Hernandez:2015ywg,Dey:2015bur,Pelaggi:2015knk,deBlas:2015hlv,Belyaev:2015hgo,Dev:2015isx,Huang:2015rkj,Moretti:2015pbj,Patel:2015ulo,Badziak:2015zez,Chakraborty:2015gyj,Cao:2015xjz,Altmannshofer:2015xfo,Cvetic:2015vit,Gu:2015lxj,Allanach:2015ixl,Davoudiasl:2015cuo,Craig:2015lra,Das:2015enc,Cheung:2015cug,Liu:2015yec,Zhang:2015uuo,Casas:2015blx,Hall:2015xds,Han:2015yjk,Park:2015ysf,Salvio:2015jgu,Chway:2015lzg,Li:2015jwd,Son:2015vfl,Tang:2015eko,An:2015cgp,Cao:2015apa,Wang:2015omi,Cai:2015hzc,Cao:2015scs,Kim:2015xyn,Gao:2015igz,Chao:2015nac,Bi:2015lcf,Goertz:2015nkp,Anchordoqui:2015jxc,Dev:2015vjd,Bizot:2015qqo,Ibanez:2015uok,Chiang:2015tqz,Kang:2015roj,Hamada:2015skp,Huang:2015svl,Kanemura:2015bli,Kanemura:2015vcb,Low:2015qho,Hernandez:2015hrt,Jiang:2015oms,Kaneta:2015qpf,Marzola:2015xbh,Ma:2015xmf,Dasgupta:2015pbr,Jung:2015etr,Potter:2016psi,Palti:2016kew,Nomura:2016fzs,Han:2016bus,Ko:2016lai,Ghorbani:2016jdq,Palle:2015vch,Danielsson:2016nyy,Chao:2016mtn,Csaki:2016raa,Karozas:2016hcp,Hernandez:2016rbi,Modak:2016ung,Dutta:2016jqn,Deppisch:2016scs,Ito:2016zkz,Zhang:2016pip,Berlin:2016hqw,Ma:2016qvn,Bhattacharya:2016lyg,D'Eramo:2016mgv}.  This has been claimed by both the ATLAS and CMS experiments with the former reporting a significance of up to $3.9\sigma$ locally and $2.3\sigma$ globally \cite{ATLAS}. The global significance represents the statistical preference for a resonance-like signal over the background, incorporating the fact that $\emph{a priori}$ the resonance could have appeared at any value of $m_{\gamma \gamma}$, a correction known as the Look-Elsewhere Effect.

In order to give a quantitative statement regarding the preference of their data for a resonance-like feature, the ATLAS collaboration must assume a functional form for their continuum background. The significance of any potential
signal then depends crucially on how well this choice was made, and whether it fully captures the uncertainties in the background near the potential resonance. This is particularly important in the case of a potential resonance at 750 GeV since this is located at an invariant mass where there is not much photon data at higher energies.  Because of this, one cannot fit the continuum background as well as in the ideal case where one has reliable data either side of the signal region where it would be possible to unambiguously determine the form of the continuum background across the signal region. Instead, one is forced to fit with the low bins and to some degree extrapolate the function to higher bins.

We seek to understand the motivation for the choice of the continuum background function made by the ATLAS collaboration in their 13~TeV diphoton analysis, and whether their choice introduced a bias into their analysis. Specifically in this article we quantify to what extent the choice of empirical function, used to model the continuum background, affects the significance of a resonance-like feature around $m_{\gamma \gamma} \sim 750$~GeV. To do this we repeat the analysis performed by the ATLAS collaboration using their form for the empirical background function and our own extension of this function.

\section{The importance of background uncertainties in the analysis}
\begin{figure*}[tb]
\centering
\includegraphics[trim={3cm 0 0 0},clip,width=0.98\textwidth]{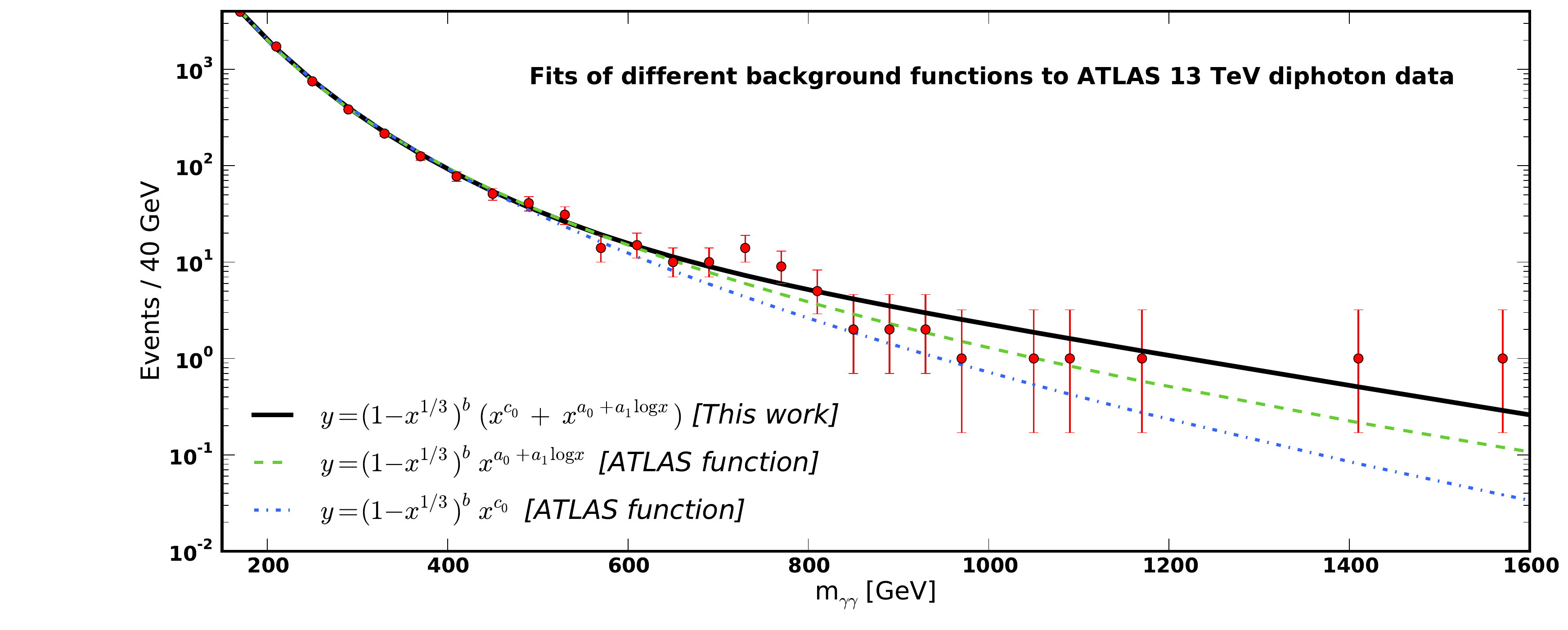}
\caption{Comparison of empirical functions for the continuum diphoton background which best fit to the data. We consider the fit function used by the ATLAS collaboration with and without the log $x$ exponent, and also
our own empirical function.}
\label{fig:funcs_plot}
\end{figure*}

The degree to which the latest ATLAS diphoton dataset prefers the presence of a resonance-like feature around 750~GeV depends crucially on our understanding of the continuum background and its uncertainties, particularly in the `signal region'
i.e. values of $m_{\gamma \gamma}$ near the potential resonance. In order to quantify this preference we employ a profile likelihood analysis, which incorporates the uncertainties on both the signal and background distributions
as nuisance parameters. 

This is done by evaluating the likelihood function for a wide range of values for the nuisance parameters, denoted by the symbol $\nu$, and finding the value of the likelihood which is largest over this range. We do this for both the background-only
scenario, where there is no signal, and where we introduce a resonance-like signal component. Following the ATLAS collaboration we write the former as $\mathcal{L}(\sigma = 0, \hat{\hat{\nu}})$ and the latter as  $\mathcal{L}(\sigma, m_X, \alpha, {\hat{\nu}})$, where $\hat{\hat{\nu}}$ denotes the values of the nuisance parameters which maximise the background-only
likelihood, and ${\hat{\nu}}$ means the same but for each non-zero value of the signal amplitude $\sigma$ (and also the central value $m_X$ and width parameter $\alpha$). The preference for a signal over background is then quantified using the test statistic, 
\begin{equation}
q_{\sigma} = -2 \, \mathrm{Log} \left[ \frac{\mathcal{L}(\sigma, m_X, \alpha, {\hat{\nu}})}{\mathcal{L}(\sigma = 0, \hat{\hat{\nu}})} \right] ,
\label{eqn:q_sigma}
\end{equation}
where we quote all logarithms to the exponential base in this work. The larger the value of $q_{\sigma}$ the greater the statistical preference for a signal feature in the data, compared to fitting with the background alone.

To simplify this process the ATLAS collaboration model the background with an empirical function, chosen to resemble the spectrum from Monte Carlo simulations. This function takes the form,
\begin{equation}
y = (1 - x^{1/3})^b x^{a_0 + a_1 \mathrm{log} x} ,
\label{eqn:atlas_fit}
\end{equation}
where $x = m_{\gamma \gamma} / \sqrt{s}$ and in their analysis the ATLAS collaboration set $a_1 = 0$. 
In the case of the profile likelihood we therefore have that for the nuisance parameters of the background $\nu = (b,a_0)$ when $a_1 = 0$ and $\nu = (b,a_0,a_1)$ otherwise. The ATLAS collaboration used a Fisher test to justify their
choice of function and also setting $a_1 = 0$, since they were able to fit the background adequately with only two parameters. However we show in section~\ref{sec:disscussion} that this conclusion is not correct, as their function
fits almost entirely to the low-energy points, which have the smallest error bars, while saying little about the region around $750$~GeV. Hence their functional choice essentially underfits the signal region and so does not capture the full background uncertainties.
In figure~\ref{fig:funcs_plot} we plot this function with $a_1 = 0$ and when allowing $a_1$ to vary freely for the best-fit parameters to the diphoton data.

Modelling the uncertainties in the background is then reduced to scanning over the parameters of the empirical function and treating them as the nuisance parameters in the profile likelihood. However this is only accurate if the variability of the empirical function within its parameter ranges is close to that of the background itself, otherwise
the analysis will be biased. Hence in full generality one should include also the uncertainty introduced through \emph{the choice of function itself}.

To understand how important this is for the ATLAS diphoton analysis, and whether it has been accounted for properly, we need to choose another suitable function to model the background. This function needs to be suitably different from equation (\ref{eqn:atlas_fit})
but must also resemble as close as possible the result from Monte Carlo simulations. 
A logical choice for a new background empirical function is to extend equation~(\ref{eqn:atlas_fit})
to a function with two components, allowing the fit near the resonance to have more freedom. Our choice therefore is a function of the form,
\begin{equation}
y = (1 - x^{1/3})^b (x^{c_0} + \, x^{a_0 + a_1 \mathrm{log} x}) .
\label{eqn:our_fit}
\end{equation}
The two-component nature of our function means that it avoids the problem whereby the low-energy data-points, which have smaller error bars, control the fit of the background
in the signal region (see section~\ref{sec:disscussion}). Here for the background-only fit we have that $\nu = (a_0,a_1,b,c_0)$.

If the resulting significance of any resonance-like feature
differs substantially when using this new function as compared to the one used by the ATLAS collaboration in their fit, then this implies that the latter function does not adequately capture the full uncertainty of the background model.
As can be seen from figure~\ref{fig:funcs_plot} our choice of function prefers larger values in the region around $m_{\gamma \gamma} \sim 750$~GeV than for the function used by ATLAS, especially compared
to the case where $c_1 = 0$. However in order to understand what effect this has on any preference for a resonance we perform a full profile likelihood analysis in the next section, as the best-fit parameters will change
with a non-zero signal contribution.

In order to confirm the suitability of the above empirical functions we have also run our own Monte Carlo simulations. We study 13 TeV proton proton collisions with two types of final states, $\gamma\gamma$ and jet + $\gamma\gamma$, obtaining the relevant processes using \textsc{MadGraph}
\cite{madgraph}. The diagrams from \textsc{MadGraph} are then passed on to \textsc{Pythia} \cite{pythia8,pythia6} for event generation and showering, using the NNPDF2.3 parton distribution function.
We then apply the ATLAS diphoton cuts, following closely the event selection procedure in \cite{ATLAS}. Both photons must satisfy $|\eta| < 2.37$  and have a minimum transverse energy $E_T > 25$ GeV. There are additional mass dependent cuts $E_T^{\gamma_1} > 0.4 m_{\gamma \gamma}$ and $E_T^{\gamma_2} > 0.3 m_{\gamma \gamma}$ where $\gamma_1$ is the photon with the greatest $E_T$, and $\gamma_2$ is the photon with the next highest $E_T$. There is a final isolation cut on each photon, $E_T^{\text{iso}} < 0.05 E_T^{\gamma} + 6$ GeV, where $E_T^{\text{iso}}$ is defined as the magnitude of the vector sum of the transverse momenta of all stable particles, excluding muons and neutrinos, in a cone of radius $\Delta R = \sqrt{(\Delta \eta)^2 +(\Delta \phi)^2} = 0.4$.
These Monte Carlo events are binned and scaled to find the expected number of events for a 13 TeV proton collider with $3.2 \rm{fb^{-1}}$ of data, which is,
\begin{equation}
	N_{\text{exp}}=3.2 \rm{fb^{-1}}*\frac{\rm{A}\sigma_{diphoton} N_{bin}}{N_{total}},
    \label{eqn:bin_scaling}
\end{equation}
where A is the acceptance ratio for standard model events given the cuts, and $\sigma_{\text{diphoton}}$ is the cross section, in fb, calculated by \textsc{Pythia} for the processes generated in \textsc{MadGraph}.
We have confirmed that both forms of the empirical function chosen by the ATLAS collaboration fit well, as does our own function. However the results of the simulation are not precise enough to prefer any particular functional dependence.  The simplest function with only $b$ and $a_0$ is a perfectly adequate fit to the simulated data, but we note that the mock data is not a perfect fit to the lower energy event rates reported in the ATLAS diphoton results. Using this simulated data to motivate the choice of background function for the real data would therefore be dangerous.

\section{Results for different background functions}
\begin{figure}[tb]
\centering
\includegraphics[width=0.49\textwidth]{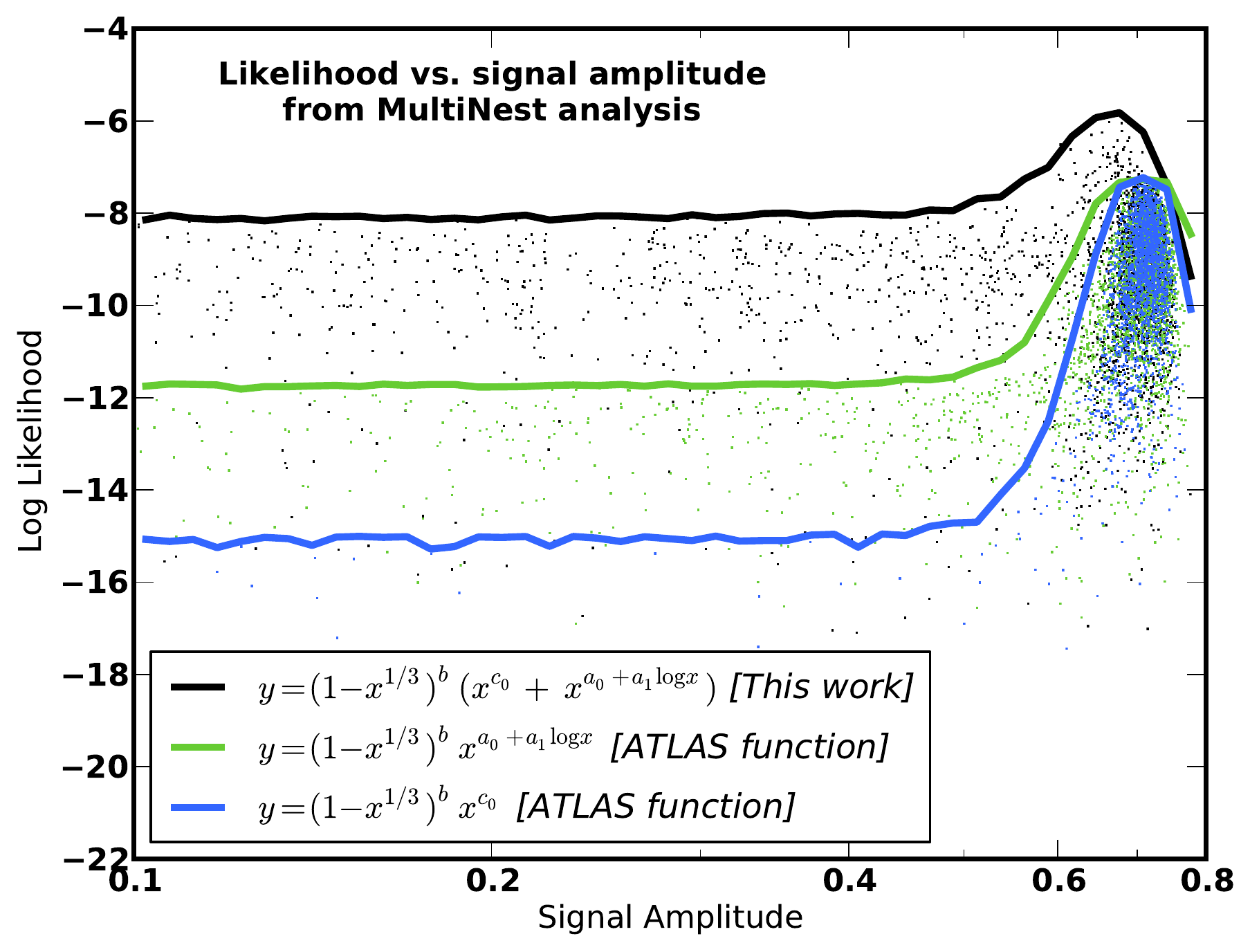}
\caption{Values of the likelihood resulting from an analysis peformed using \textsc{MultiNest} plotted against the amplitude of a potential resonance added to one of three different choices for the continuum diphoton background. The lines give the maximum likelihood for each amplitude value.  'Signal Amplitude' refers to the prefactor multiplying the signal contribution, which is a resonance normalised to unity with mass between 700 and 800 GeV.}
\label{fig:likes_all}
\end{figure}

In the previous section we discussed how the modelling of the smooth component of the background with an empirical function, for a search for potential resonances in the diphoton data, is complicated by the need to incorporate not only the uncertainties 
in the empirical function itself, but also in the choice of function. We seek to understand if this was adequately accounted for in the analysis of the ATLAS collaboration.

We show in figure~\ref{fig:likes_all} the amplitude of a resonance-like feature compared with the value of the likelihood function, resulting from a profile likelihood analysis performed using \textsc{MultiNest} \cite{multinest1,multinest2,multinest3},
when this feature is added to each of the background empirical functions. Each dot represents a particular set of nuisance parameters, while the lines mark the maximum likelihood for each amplitude value, used for the profile likelihood analysis. 
The relative height of the peak compared to the likelihood as the amplitude tends to zero then gives a result proportional to $q_{\sigma}$.
We use a Breit-Wigner distribution to model the shape of the resonance-like feature, however our results have been cross-checked using a Crystal Ball distribution instead~\cite{ATLAS}. When using the ATLAS background function i.e. equation~(\ref{eqn:atlas_fit}) with $a_1 = 0$ there is a clear preference for a resonance-like feature around 750~GeV, in agreement with the results from the analysis of the ATLAS collaboration. Indeed the significance of this preference is $3.9\sigma$ when allowing the width of the resonance to vary freely, as detailed also in table~\ref{table_sigmas}. 
For the Narrow Width Approximation (NWA) we assume a Crystal Ball function for the signal with a width fixed to the photon energy resolution~\cite{ATLAS}.

However when allowing $a_1$ to vary and treating it as an additional nuisance parameter in the profile likelihood, the preference for such a feature drops to $2.9\sigma$ with a freely varying width. The effect is even more drastic when using
our own empirical function i.e. equation~(\ref{eqn:our_fit}), where the preference for a resonance is now much lower at approximately $2\sigma$ local significance. The fit now also prefers a smaller resonance, as expected since
the best-fit form of this function prefers a larger continuum background in the signal region (see figure~\ref{fig:funcs_plot}).

Hence it is clear that the significance of a potential resonance-like feature depends strongly on the choice of empirical function used to model the background, as summarised in table~\ref{table_sigmas}. 
Indeed the sensitivity of the analysis to a change in the continuum background function causes severe concern, and places doubt on the statistical significance of this feature in the ATLAS 13~TeV diphoton data.
In the next section we address the issue of free parameters in the background function, and show that the choice made by ATLAS is underfit in the region around 750 GeV.

\begin{table}[t]
\begin{center}
\begin{tabular}{ c || c | c  }
\normalsize{Background function} & \normalsize{Free width} & \normalsize{NWA} \\
\hline
\normalsize{$y = (1 - x^{1/3})^b x^{a_0}$} & \normalsize{$3.9 \sigma$} & \normalsize{$3.6 \sigma$} \\
 \normalsize{$y = (1 - x^{1/3})^b x^{a_0 + a_1 \mathrm{log} x}$} & \normalsize{$2.9 \sigma$} & \normalsize{$2.6 \sigma$}  \\
  \normalsize{$y = (1 - x^{1/3})^b (x^{c_0} + \, x^{a_0 + a_1 \mathrm{log} x})$} & \normalsize{$2.0 \sigma$} & \normalsize{$2.0 \sigma$} \\
\end{tabular}
\end{center}
\caption{Local significance for a resonance-like signal at $m_{\gamma \gamma} \sim 750$~GeV under different assumptions for the functional dependence of the smooth background. The first function is the one used by ATLAS in their 
analysis. We either allow the width of the resonance to vary freely, or keep it fixed in the case of the Narrow Width Approximation (NWA).}
\label{table_sigmas}
\end{table}

\section{Discussion of results \label{sec:disscussion}}
\begin{figure}[b]
\includegraphics[width=0.49\textwidth]{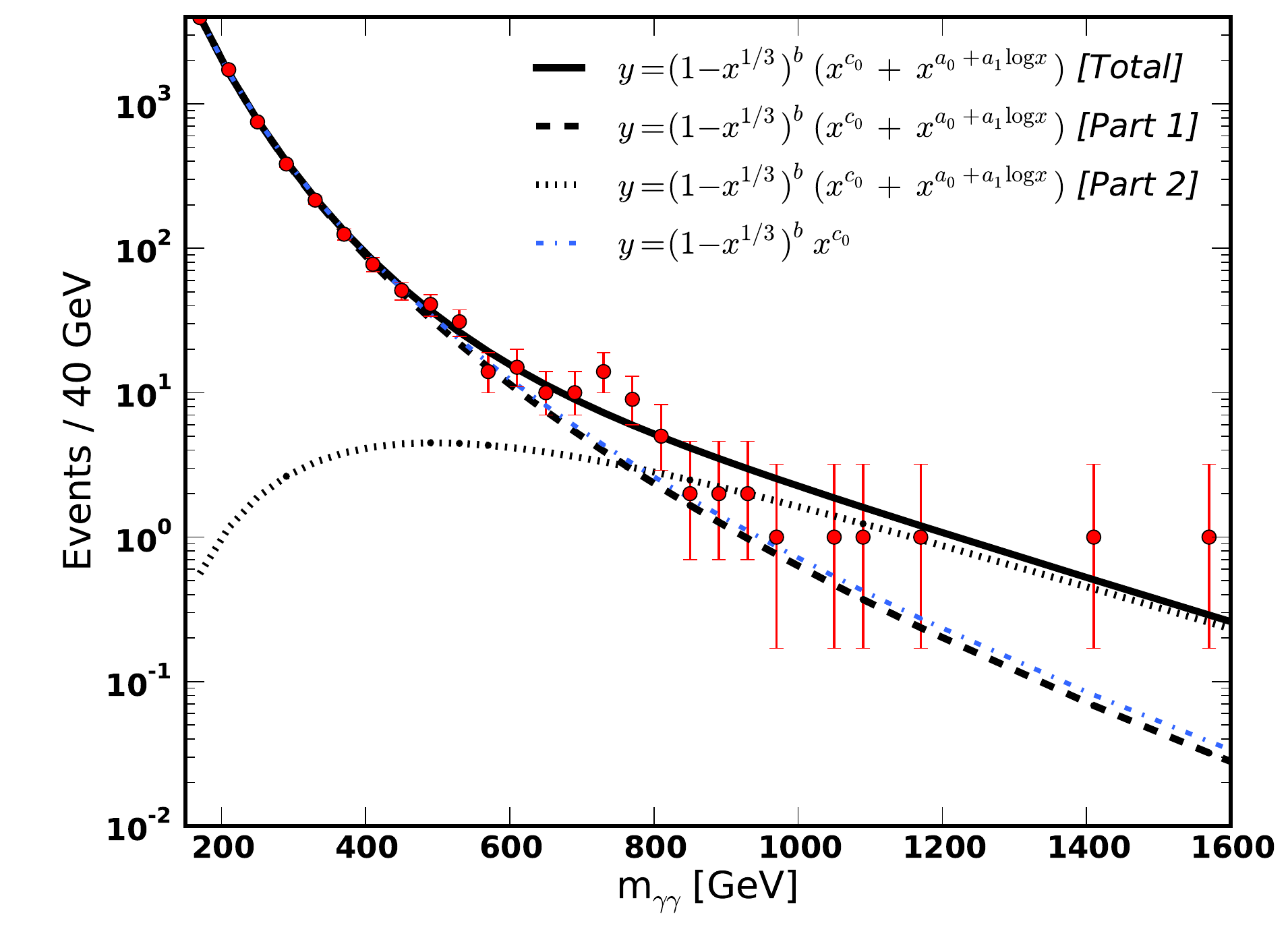}
\caption{Comparison of the best-fit form of the function used by the ATLAS collaboration in their own analysis i.e. equation~(\ref{eqn:atlas_fit}) with both components of our own best-fit function i.e. equation~(\ref{eqn:our_fit})
labelled as `Part 1' and `Part 2' (and the sum of these two as the `Total'), and the diphoton data from the ATLAS 13~TeV run.}
\label{fig:bg_func_parts}
\end{figure}

We have found that the significance of a resonance-like feature in the 13 TeV ATLAS diphoton data around an invariant mass of 750~GeV depends sensitively on the choice of function used to model the continuum background. In this section
we seek to understand why the analysis performed by the ATLAS collaboration may have over-estimated the significance of a signal feature. 

An important issue is the number of parameters on which the background function depends \emph{in the signal region}. In their own analysis~\cite{ATLAS} the ATLAS Collaboration justified their rather simple two-parameter function (equation~(\ref{eqn:atlas_fit})
with $a_1 = 0$) using a Fisher test, which showed that adding an additional free parameter (i.e. allowing $a_1$ to take on any value) did not improve the fit to the diphoton data enough to justify its inclusion. However here we show
that their choice of function is fit almost entirely to the low-energy region. Hence the result of the Fisher test performed by ATLAS has little relevance for the background in the region around $750$~GeV.

We illustrate this with figure~\ref{fig:bg_func_parts}, where we show both components of our function i.e. equation~(\ref{eqn:our_fit}) labelled as `Part 1' and `Part 2', compared to the function the ATLAS collaboration use in their analysis i.e. equation~(\ref{eqn:atlas_fit})
with $a_1 = 0$. The important point to note is that the first component of our best-fit function, obtained by fitting to the data without any signal component, is almost identical to the form of the ATLAS best-fit function.
This implies that the latter is determined almost entirely by the low-energy points, as expected, while its form at high energy near the potential resonance does not depend strongly on the data in this region.
Hence the function is at best under-fitting to the background in the signal region, at at worst hardly fitting to the data in this region at all.
Indeed the fit of our function in figure~\ref{fig:bg_func_parts} shows clearly that the low and high energy regions (below and above $\sim 500$~GeV) are in significant tension, since they prefer different background spectra. 

The issue is exacerbated for this particular data-set due to the lack of data at values of $m_{\gamma \gamma}$ larger than $\sim 750$~GeV. This is because the background function is only effectively fixed at the low-energy
end, while its value at higher energies has much more freedom. If instead there were much more data at energies above $\sim 750$~GeV then it is unlikely the choice of background function would make much difference to the final
result of the profile likelihood analysis, since all fits would then give the same continuum fit in the signal region.

The Fisher test is not the only method of estimating the required number of free parameters. An alternative metric is the Bayesian Information Criterion (BIC) \cite{BIC}, which takes the form $\mathrm{BIC} = -2\ln\mathcal{L} + k \ln n$, where $k$ is the
number of parameters in the model and $n = 27$ is the number of data-points. The model with the lowest value of BIC is the best choice, however only if the difference $\Delta \mathrm{BIC} \gtrsim 2$, otherwise both models provide equally good fits.
For the simplified ATLAS model with $a_1 = 0$ we have that $k = 2$ and so $\mathrm{BIC} \approx 36.6$ without any signal component, while when $a_1 \neq 0$ we
find $\mathrm{BIC} \approx 33.4$ with $k = 3$ and for our own function (equation~(\ref{eqn:our_fit})) we have $k = 4$ and so $\mathrm{BIC} \approx 29.2$. Hence under the BIC our model for the background is justified despite its increased
number of free parameters, as it clearly fits the background better over the whole range of invariant masses.

Of course fitting a resonance to the data with the simplest background function only has one more degree of freedom compared to our two extra parameter background function, but the change in $\mathrm{BIC}$ is less favourable in that case. Even if this was not the case, only when the background-only hypothesis becomes severely disfavoured could we be sure that a new resonance has appeared in teh data.  It goes without saying that we very much hope that such a resonance exists has and this note of caution is unnecessary.

In summary we have doubts over the application of the Fisher test by the ATLAS collaboration, which justified their choice of simplified background function, as their fits are dominated by data points at much lower invariant
mass than the tentative resonance. We showed that an alternative test, the Bayesian Information Criterion, which gives the best-fit model weighted by its number of free parameters, has a clear preference for our own empirical model
over either of the ATLAS functions even given its additional degrees of freedom.

\section{Conclusion}
The first data-release from collisions of protons with a centre-of-mass energy of $\sqrt{s} = 13$~TeV at the LHC has lead to claims from both the ATLAS and CMS experiments of a preference in their data from the diphoton channel for a resonance-like
feature around a centre-of-mass energy of $m_{\gamma \gamma} \sim 750$~GeV.
Indeed the analysis performed by the ATLAS collaboration finds at most a $3.9\sigma$ local preference for a resonance-like feature in their diphoton data around 750~GeV~\cite{ATLAS}.

The ATLAS analysis was performed by making an assumption on the form of the continuum background for this search, based on knowledge from Monte Carlo simulations. The significance of a resonance in the data above the background therefore
depends crucially on how well this empirical function captures the uncertainties of the background, especially in the region of $m_{\gamma \gamma}$ where the resonance is claimed to be present, and where there is little data compared
to lower energies. 

In this work we have quantified to what extent the preference of the data for a resonance-like feature around 750~GeV depends on the choice of this empirical function. To do this we have written down a new function which allows the background around $750$~GeV to be fit independently of the low-energy region (see figure~\ref{fig:funcs_plot}). 

By performing a profile likelihood analysis using our own empirical background function and the one use by the ATLAS collaboration we have calculated the significance of a resonance-like feature around $750$~GeV. The results of this
analysis are shown in figure~\ref{fig:likes_all} and table~\ref{table_sigmas}. We find that the results of the analysis are highly sensitive to the choice of background function, and that if we use our own form the preference for
a resonance-like feature is only at the level of $2 \sigma$ locally. 

The reason for this disagreement with the analysis of the ATLAS collaboration~\cite{ATLAS} is that their choice of function fits almost entirely to the data at low-energy, while underfitting (or extrapolating) in the region around
$750$~GeV. Hence while they found, using a Fisher test, that only two free parameters were needed to adequately describe the full continuum background, this was only for the low energy region region, and not for diphoton invariant mass near the 
potential resonance. We showed in figure~\ref{fig:bg_func_parts} that an additional
component is needed in the function to describe the region above $500$~GeV, and to fully capture the uncertainties in the background around the potential resonance. 
Additionally the fact the Bayesian Information Criteron gives a clear preference for our own model despite its increased number of free parameters is suggestive that the background function used by the ATLAS collaboration is not adequate.  

We attempted to model the standard model diphoton background and, like the ATLAS collaboration, found no evidence requiring a more complicated fit than that of equation (\ref{eqn:atlas_fit}) with $a_1=0$.  However, we understand (and have found both in this situation and in others) that the precise modelling of LHC background events with Monte Carlo generators is very challenging - the ATLAS collaboration do not use their Monte Carlo simulations to fit the background events but rather to motivate the functional form of the fit to the background in the data.  If the explanation of the deviation between this simple curve and the data is due to an incorrectly chosen functional form for the background, this may point to new insight into standard model physics.  

In summary we have found that the background is not known well enough, and the high-energy data not yet precise enough, to make such a strong statement on the presence of such a resonance. Instead we find a local significance for such a feature at the level of only $2 \sigma$ in the ATLAS 13~TeV diphoton data if we assume a rather simple extension of the background model.

The fact that different statistical treatments might lead to different interpretations clearly indicates the need for more data.  We hope to see the next run of the LHC provide this data and continue its groundbreaking test of high energy physics.
%RIP DAVID BOWIE 1947-2016

\section*{Acknowledgements}
The research leading to these results has received funding from the European Research Council through the project DARK HORIZONS under the European Union's Horizon 2020 program (ERC Grant Agreement no.648680).  JH and MF are also grateful for funding from the UK Science and Technology Facilities Council (STFC).

%\bibliography{diphoton_refs}{}
%\bibliographystyle{kp.bst}
%\end{document}

\begingroup\raggedright\endgroup

\end{document}